\documentstyle[preprint,eqsecnum,aps,prb]{revtex}

\begin{document}
\title{Colossal magnetoresistive manganite thin films}
\author{W. Prellier\thanks{%
email: prellier@ismra.fr}, Ph. Lecoeur and B.\ Mercey}
\address{Laboratoire CRISMAT, CNRS\ UMR 6508, Bd du\\
Mar\'{e}chal Juin, 14050 Caen Cedex, FRANCE.}
\date{\today}
\maketitle

\begin{abstract}
Mixed-valence perovskite manganites (Re$_{1-x}$A$_{x}$MnO$_{3}$ where
Re=rare earth, A=alkaline earth) provide a unique opportunity to study the
relationships between the structure and the magnetotransport properties due
to an interplay among charge carriers, magnetic coupling, orbital ordering
and structural distortion. This makes these compounds very exciting from
both the basic research and from the technological view point. As the
technology pursued with these materials requires film growth, extensive
studies have been made on materials synthesis, structural and physical
characterization and device fabrication. In this article, the results of the
different experimental techniques and the effects of the deposition
procedure of the manganite thin films are first reviewed. Second, the
relation between the structural and the physical properties mentioned, and
the influence of strains discussed.\ Finally, possible applications of
manganite thin films for spin electronics are presented.
\end{abstract}

\bigskip

\newpage

\section{Introduction}

The last decade has seen the emergence of epitaxial metal-oxide films as one
of the most attractive subjects for the condensed matter community. The
emergence of such interest was primarily stimulated by the discovery of
high\ temperature superconductors (HTSC) and more recently by the discovery
of the colossal magnetoresistance (CMR) effect in thin films of manganese
oxides Re$_{1-x}$A$_{x}$MnO$_{3}$ (where Re is a rare earth and A is an
alkaline earth) \cite
{Jin94a-Cro,Jin94b-Cro,VonHelm93-Cro,VonHelm94-Cro,McCom94-Cro}. CMR\
materials exhibit large changes in electrical resistance when an external
magnetic field is applied \cite{Tokura-Liv,Raveau-Liv}.

The doped manganites are mixed-valence with $Mn^{3+}$ ($3d^{4}$) and $%
Mn^{4+} $ ($3d^{3}$). For the octahedral site symmetry of MnO$_{6}$ the
configurations become $t_{2g}^{3}e_{g}^{1}$ for $Mn^{3+}$ and $t_{2g}^{3}$
for $Mn^{4+}$. In the double-exchange mechanism, the $e_{g}$ electrons are
considered as mobile charge carriers interactions with localized $Mn^{4+}$
(S=3/2) spins. The carriers hoping avoids the strong on-site Hund rule
exchange energy $J_{ex}$ when the spins are aligned ferromagnetically. (Note
that if the Mn spins are not parallel or if the Mn-O-Mn bond is bent, the
electron transfer becomes more difficult and mobility decreases). $J_{ex}$
is much larger that the $e_{g}$ bandwidth and thus, the conduction electrons
are highly spin-polarized in the ground state. With this idea, correlations
of the half ferromagnetic character of the CMR\ materials were found\cite
{Half}. Theoretical and experimental studies indicate that the small polaron
effects including Jahn-Teller distorsion also play important roles for the
transition and transport measurements as well\cite{Theory}.

These oxide materials are important from a fundamental point of view since
they offer a chemical flexibility that enables new structures and new
properties to be generated and, consequently, the relations between the
structure, electronic, magnetic and transport properties to be studied.

Since most technological applications require thin films on substrates, the
ability to prepare \ such films and understand their properties is of a
prime importance. The\ synthesis of the first high temperature
superconducting oxide thin films almost 15 years ago generated great
interest in the thin films community. This resulted in the development of
various techniques, guided by the importance of preparing high quality thin
films of superconductor compounds, including sputtering, Molecular Beam
Epitaxy (MBE) and Metal Organic Chemical Vapor Deposition (MOCVD), but the
most popular technique is probably the Pulsed Laser Deposition (PLD)\cite
{Chrisey}. This latter method is used extensively to synthesize cuprates and
HTSCs, which are now routinely made in laboratories, and it has been easily
and rapidly adapted for manganites. Another reason for this quick-transfer
technology is that these oxide materials crystallize in a perovskite
structure as the HTSCs\ and in some sense, they are quite similar \cite
{Venky96-Liv,Venky99-Div41}. Moreover, the manganite oxides are highly
sensitive to the strain-effect, and this offers the possibility of studying
its influence upon various properties such as insulator-to-metal transition
temperature ($T_{IM}$), Curie temperature ($T_{C}$), structure,
microstructure/morphology, etc...\ The renewed interest in the manganite
materials has resulted in a large volume of published research in this
field.\ 

In the present paper, we present a brief review of the experimental work
done in the past 7 years.\ The deposition procedure and its influence
(through deposition temperature, oxygen pressure, post-annealing, substrate
type...) upon the magneto-transport properties will be discussed. In the
particular case of thin films, work was mainly devoted to substrate-induced
strain and thickness dependence, and we will describe the experimental
situation.\ Finally, since much interest in the community of the CMR thin
films lies in their use in devices, some applications of these material will
also be presented. Due to an extremely large amount of research data
published in this field, there are unfortunately some missing citations, and
we apologize the authors in advance.

\section{Deposition procedures, Structure and Properties}

CMR\ manganite materials are compounds crystallizing in a perovskite-like
structure, which apart from manganese and oxygen, contain rare earths and/or
lanthanide cations. The prototype compound is La$_{1-x}$Ca$_{x}$MnO$_{3}$,
but there are many other related structures. Numerous studies were performed
on the hole doped La$_{0.7}$A$_{0.3}$MnO$_{3}$ manganites (where A=Sr or Ca)
since these CMR\ materials exhibit so far the highest Curie temperatures
(often associated with an insulator-to-metal temperature transition). Such
high transition, close to room temperature, make them suitable for
applications \cite{Venky98-Div23}. As previously said, there are many
different compounds due to the fact that the A-site cation can be a
lanthanide or a rare earth.\ Thus, a number systems have been studied in the
form of thin films including: La-Ca-Mn-O \cite
{Jin94a-Cro,Jin94b-Cro,VonHelm93-Cro,McCom94-Cro,Lawler95-Cro,Zeng95a-Cro,Gu95-Cro,Lawler96-Cro,Gommert97-Cro,Lebedev-PRB,Prellier99-Cro,Thomas98-Div23,Guo97-Cro,Cao00-Cro,Lu00-Cro}%
, La-Sr-Mn-O \cite
{Trajanovic96-Cro,Snyder96-Cro,Izumi98-Cro,Foncuberta99b,Li99-Cro,Xiong00-Cro,Dulli00-Cro,Gonzalez00-Cro}%
, La-Ba-MnO \cite{Robson96-Book,Zhu99-Cro,Kanki00-Cro}, La-Pb-Mn-O \cite
{Manoharan94-Cro,Srinivasan95-Cro,Yamada00a-Cro,Yamada00b-Cro,Borca00-Cro},
La-Mn-O \cite{Gupta95-Cro,Chen98a-Cro,Zhao99-Sta}, Nd-Sr-Mn-O \cite
{Xiong95a-Cro,Xiong95b-Cro,Xiong95c-Cro,Xiong96-Cro,Kasai96b-CO,Wagner98-CO,Prellier99-Thick,Ponmambalam99-CO,Wang00-Cro}%
, Sm-Sr-Mn-O \cite{Kasai96a-CO,Oshima99-CO}, Pr-Ca-Mn-O \cite
{Singh96-Cro,Prellier00-APL,Haghiri00-JAP,Lee00-CO,Saraf00-CO}, Pr-Sr-Mn-O 
\cite{Mercey97-CM,Wagner97a-CO,Wagner97b-CO,Wang00-Sta,Mercey00-CM},
La-Sn-Mn-O \cite{Guo00-Cro}, La-K-Mn-O \cite{Chen97-Cro}, La-Ce-Mn-O \cite
{Zhao00-Cro} and Bi-Sr-Mn-O \cite{Ohshima00-Cro}. To detail the situation in
terms of synthesis, this section is divided into three parts.\ First, in
part A, we will review the parameters that govern the growth of these oxides
(temperature, oxygen pressure etc...), and we will consider the progress
made in the synthesis of these materials. We will also focus on the
different techniques that were are for the growth of thin films.\ Then in
part B, we will discuss the different studies that were carried out to
understand the structure and the microstructure of the thin films for the
simple perovskite oxides and also for the results of the double ordered
perovskites. Finally, some physical measurements will be presented (part C).

\subsection{Synthesis of manganite films}

\subsubsection{Simple perovskite AMnO$_{3}$}

The manganite thin films have been mainly prepared using the Pulsed Laser
Deposition technique \cite
{Jin94b-Cro,Li99-Cro,Xiong95a-Cro,Pietambaram99-Sta,Lourenco99-Cro,Ziese99-Cro}%
. The principle of this technique is relatively simple.\ A pulsed laser beam
ablates a dense ceramic target of the desired material. In the presence of a
background gas (usually oxygen), a plasma is produced and condenses on the
heated substrate. Typical lasers used for manganite are excimer UV with KrF
\ at $\lambda =248nm$ \cite{Prellier00-APL,Fontcuberta99a}, XeCl at $\lambda
=308nm$ \cite{Lu00-Cro,Singh96-Cro,Ziese99-Cro} or ArF at $\lambda =193nm$ 
\cite{Kanki00-Cro}. A frequency tripled Nd-YAG at $\lambda =355nm$ \cite
{Gu95-Cro,Gupta95-Cro} or quadrupled Nd-YAG at $\lambda =266nm$ \cite
{Yamada00a-Cro} may also be used.\ A cross-beam deposition scheme utilizing
two Nd-YAG lasers was also used to grow Pr$_{0.65}$Ca$_{0.35}$MnO$_{3}$ thin
films on LaAlO$_{3}$ \cite{Lee00-CO}. However, the utilization of high
oxygen pressure during the PLD growth\ prevents the use of Reflection High
Energy Electron Diffraction (RHEED)\ system in order to control in-situ the
different stages of the growth. By a more oxidizing gas (atomic oxygen,
ozone..) and a differential pumping system, the electron-path in the high
pressure oxygen atmosphere can be reduced, and thus the specular beam of the
RHEED can be monitored in order to observe the oscillations (see Fig. 1 for
a typical experimental setup). High quality manganite thin films were
fabricated in this way \cite{Izumi98-Cro,Zhu99-Cro,Klein00-Cro}. A
persistent intensity of the RHEED\ is observed and the roughness of the
films is low, around one unit cell \cite{Izumi98-Cro}.

A second popular deposition technique is magnetron sputtering, which can be $%
RF$ \cite
{Gonzalez00-Cro,Srinivasan95-Cro,Mercey97-CM,Dorr00-Cro,Vlakhov98-Sta} or $%
DC $ \cite{Cao00-Cro,Xiong00-Cro,Wagner97a-CO,Zeng95b-Cro,Li97-Cro}.
Reactive sputtering is particularly useful for large-area films, but the
deposition of complex oxides, comprising several cations, is difficult
because of a possible change in the material composition between the target
and the film. Regardless, in these two techniques (PLD and sputtering)
utilize highly-dense ceramic targets, and the configuration is usually
''on-axis''.\ This means that the plane of the substrate is perpendicular to
the particles flux. Samples can\ also be produced by sputtering \cite
{Wang00-Cro,Cheng98-Cro,Broussard99-Cro} and PLD \cite
{VonHelm93-Cro,Gommert97-Cro} in the ''off-axis''configuration .\ In the
case of PLD, this decreases the surface roughness and avoids the formation
of droplets associated with laser deposition. Ion beam sputtering \cite
{Chahara93-Cro}, electron beam/thermal coevaporation \cite{Chen97-Cro} and
molecular beam epitaxy \cite{O'Donnel96-Anis,Gross00-Cro,Reutler00-Cro} have
also been utilized to make manganite thin films. MOCVD was used to prepare
high quality thin films of various compositions \cite
{Snyder96-Cro,Pignard98-Cro,Dubourdieu99-Cro,Zhu00-Cro}. Briefly,
metal-organic precursors are dissolved in an ether and injected into the low
pressure apparatus. In contrast to the methods previously described, this
one does not require a high vacuum but allows deposition at higher oxygen
pressure. Others methods used for making manganite thin films are nebulized
spray pyrolysis \cite{Ponmambalam99-CO,Parashar00-CO} and sol-gel \cite
{Bae96-Cro}.

Various gases, such as O$_{2}$ \cite{Prellier00-APL,Ziese98-Anis}, N$_{2}$O 
\cite{Xiong95a-Cro,Xiong95b-Cro}, ozone \cite{O'Donnel00-Anis} or a mixture
of argon-oxygen atmosphere \cite
{Mercey97-CM,Broussard99-Cro,Steenbeck99-Anis}, result in
oxygen-stoichiometric films. The background gas pressure is important for
the oxidation process. Thus, the emission of the gas-phase oxidation of Mn
during pulsed laser deposition of manganites in O$_{2}$ and N$_{2}$O
atmospheres was studied. They shown that both oxidation in the gas phase and
at the surface are required in order to obtain the optimized properties. It
was found that N$_{2}$O increases the oxidation of Mn in the plasma plume,
leading to an improvement of the magnetic properties of La$_{0.67}$Sr$%
_{0.33} $MnO$_{3}$\cite{Lecoeur96-Cro}.

The deposition conditions (oxygen pressure $P_{O2}$, deposition temperature $%
T_{S}$, laser fluence...) can drastically influence the properties\cite
{Gu95-Cro,Guo97-Cro,Mercey00-CM,Rajeswari98-Div20,Liu00-Cro}.\ For Nd$_{0.7}$%
Sr$_{0.3}$MnO$_{3}$ grown on (100)-LaAlO$_{3}$ \cite{Xiong95b-Cro}, the
maximum resistivity peak\ shifts to lower temperatures as the deposition
temperature decreases (the optimum $T_{C}$\ of $175K$ is obtained for $%
T_{S}=615%
{{}^\circ}%
C$). Yamada {\it et al. }\cite{Yamada00a-Cro} have shown that the $T_{IM}$
decreases with either $T_{S}$ or $P_{O2}$ in La$_{1-x}$Pb$_{x}$MnO$_{3}$.\
The deposition temperature also strongly influences the microstructure of Pr$%
_{0.7}$Sr$_{0.3}$MnO$_{3}$, since films grown at low temperature exhibit a
columnar growth with well-connected grains while those deposited at higher
temperature are poorly connected with platelet-like crystals \cite
{Mercey00-CM}.

In addition, it has also been shown that in-situ \cite{Gu95-Cro,Wu00-Cro} or
ex-situ \cite{Guo97-Cro,Dulli00-Cro,Gonzalez00-Cro,Lobad00-Cro,Nam01-Cro}
oxygen annealings is necessary to obtain the optimized properties. In
particular, the postannealing of the films can lead to significant
modification of the oxygen content and optimizes the physical properties 
\cite{Li99-Cro,Ju94-Liv,Pignard97-Cro} such as $T_{IM}$ \cite{Xiong95c-Cro}, 
$T_{C}$ and the CMR\ effect \cite{Jin94a-Cro,VonHelm93-Cro,McCom94-Cro}.
This annealing effect is necessary to achieve the optimum oxygen
concentration of the films. Depending on the nature of the film and on the
growth conditions, annealing is used either to fully oxidize the film
(oxidative annealing) or to remove extra oxygen (reductive annealing). The
effect of annealing was also observed in films annealed in N$_{2}$
atmosphere (see Fig. 2 for La$_{1-x}$Sr$_{x}$MnO$_{3}$) \cite{Ju94-Liv}. The 
$T_{IM}$\ transition shifts after the annealing to higher temperature and
the MR ratio increases slightly. The magnetic transition also occurs at
higher temperature after annealing. For example, the Curie temperature, $%
T_{C}$, is found to increase from 200K for as-deposited La$_{0.8}$MnO$_{3}$
film to 320K after the fourth thermal treatment \ (See Fig. 3) \cite
{Pignard97-Cro}. In fact, the changes under annealing\ can also be seen on
the position of the diffraction peak \cite{Prellier00-APL,Pignard97-Cro}:
the out-of-plane parameter decreases, which relates to an increase of the $%
Mn^{3+}/Mn^{4+}$ ratio \cite{Prellier99-Cro,Pignard97-Cro}.\ This annealing
effect can lead to an improvement in physical properties (like in La$_{0.8}$%
Ca$_{0.2}$MnO$_{3}$ where $T_{IM}$ value is higher than that of the bulk\cite
{Shreekaka99a-Cro}) but sometimes the resistivity peak is lower by more than
100K \cite{Choi01-Cro}. In fact, Prellier {\it et al.} have shown that the
entire phase diagram is different in the ferromagnetic region of La$_{1-x}$Ca%
$_{x}$MnO$_{3}$ ($0<x<0.5$) \cite{Prellier99-Cro} as compared to the bulk
(in terms of transport and magnetic transitions).

Doping is another method used to improve the magnetic properties. For
example, enhancement of the properties ($T_{IM}$\ and $T_{C}$) is observed
with silver addition to the La$_{0.7}$Ca$_{0.3}$MnO$_{3}$ target ($5\%wt.$) 
\cite{Shreekala99b-Cro} or with La$_{2/3}$Sr$_{1/3}$MnO$_{3}$ films doped
with Ag and grown by dual-beam PLD \cite{Li01-Cro}.

\subsubsection{Double and triple perovskites: A$_{3}$Mn$_{2}$O$_{7}$ and A$%
_{4}$Mn$_{3}$O$_{10}$}

Although the majority of studies have been done on simple perovskites AMnO$%
_{3}$, colossal magnetoresistance also occurs in (La,A)$_{3}$Mn$_{2}$O$_{7}$
(A=Ca,Sr). These compounds belong to the Ruddlesden-Popper phases whose
general formula is A$_{n+1}$B$_{n}$O$_{3n+1}$. Two parents of this family
were synthesized in thin film form with $n=2$ and $n=3$.

Films of La$_{2-2x}$Ca$_{1+2x}$Mn$_{2}$O$_{7}$ ($x=0.3$) were deposited on
(001)-MgO by single-target magnetron sputtering \cite
{Asano97a-Cro,Asano97b-Cro}. $c$-axis oriented films of La$_{2-2x}$Sr$%
_{1+2x} $Mn$_{2}$O$_{7}$ ($x=0.4$) can be grown on (001)-SrTiO$_{3}$ under
limited conditions (above 900$%
{{}^\circ}%
$C for the deposition temperature and below $100mTorr$ for the oxygen
partial pressure, Fig. 4)\cite{Konishi98-Cro} which are different from the
typical conditions used for (La,Sr)MnO$_{3}$ films. On SrTiO$_{3}$
substrates, the resistivity curves shows a transition at $100K$ which
coincides with a magnetic transition for La$_{2-2x}$Sr$_{1+2x}$Mn$_{2}$O$%
_{7} $\ films on SrTiO$_{3}$ \cite{Konishi98-Cro}. Films on MgO are $a$-axis
oriented, which means that the long parameter is in the plane of the
substrate, these films evidence two types of ferromagnetic ordering that
possibly result from anisotropic exchange interactions for $0.22<x<0.55$ 
\cite{Asano97-PRB}. Magnetoresistance is observed in a wide temperature
range \ below the ferromagnetic transitions on MgO and is accompanied by an
hysteresis on SrTiO$_{3}$ \cite{Konishi98-Cro}. Epitaxial films of (La,Sr)$%
_{3}$Mn$_{2}$O$_{7}$ can also be grown artificially by atomic-layer stacking
of SrO and (La,Sr)MnO$_{3}$ \cite{Tanaka00c-SL}.

Contrary to the previous one, the La$_{3-3x}$Ca$_{1+3x}$Mn$_{3}$O$_{9}$ ($%
x=0.3$) compound can be stabilized, but only in the form of thin films and
not in the form of bulk\cite{Asano97c-Cro}.\ Features similar to those
reported for the double perovskite ($n=2$) were also observed for the $n=3$
compound indicating a correlation between the dimensionality (or the $c$%
-axis bond configuration) and the magneto-transport properties \cite
{Asano97c-Cro}.

Thus, it appears that the increasing of the c-axis reduces the magnitude of
the of the CMR at low temperature and this may be attributed to the increase
magnitude of the double-exchange tranfer matric and a better ferromagnetic
spin alignement.

\subsubsection{The particular case of the ordered double perovskite Sr$_{2}$%
FeMoO$_{6}$}

This review focuses on manganite thin films, but it is also interesting to
present the results on the ordered double perovskite Sr$_{2}$FeMoO$_{6}$ ,
even though it does not contain Mn, since it exhibits magnetoresistance \cite
{Kobayashi98} with a Curie temperature above $370K$. Films of Sr$_{2}$FeMoO$%
_{6-y}$ were grown, using pulsed laser deposition on (001)-SrTiO$_{3}$ \cite
{Asano99-Cro,Westerburg00a-Cro}. They are grown on both (001) and (111)-SrTiO%
$_{3}$ but in a narrow window near $900%
{{}^\circ}%
C$ and $10^{-6}Torr$ \cite{Manako99-Cro}. Asano{\it \ et al.} \cite
{Asano99-Cro} have shown that by altering the growth conditions they are
able to induce either positive ($35\%$ ) or negative ($-3\%$)
magnetoresistance at $5K$ under a magnetic field of $8T$.\ The films show
metallic conductivity with a ferromagnetic transition above $400K$ \cite
{Manako99-Cro}. The experimental magnetic moment is calculated to be $4\mu
_{B}$ per formula unit \cite{Westerburg00a-Cro} in agreement with the
theoretical one\cite{Kobayashi98}. Sr$_{2}$FeMoO$_{6}$ films also exhibit
both an electron-like ordinary Hall effect and a hole-like anomalous Hall
contribution \cite{Westerburg00b-Cro}. More importantly, an intergrain
tunneling type low field magnetoresistance\cite{Yin99-Cro}, even at room
temperature\cite{Manako99-Cro,Yin00-Cro}, has been reported.

\subsection{Structure and microstructure}

It is of prime importance to carry out structural characterization of the
films, since it has been shown that in the bulk material, a slight variation
of the Mn-O bond length or bond angle drastically modifies the physical
properties. Consequently, careful characterization of AMnO$_{3}$ films is
paramount especially\ from the crystal structure point of view\cite
{Prellier01-Com}.

One of the best techniques to study the local structure of thin films, as
for bulk, is most probably the high resolution transmission electron
microscopy (HREM). Van Tendeloo {\it et al. }\cite{VanTendeloo00-Stu} have
studied the evolution\ of the microstructure as function of the thickness in
La$_{0.7}$Sr$_{0.3}$MnO$_{3}$ films on LaAlO$_{3}$.\ Close to the interface,
both the film and the substrate are elastically strained in opposite
directions in such a way that the interface is perfectly coherent.\ In the
thicker films, the stress is partly relieved after annealing by the
formation of misfit dislocations. Similar results were found for La$_{1-x}$Ca%
$_{x}$MnO$_{3}$\cite{Lebedev-PRB}, where the bottom part of the film, close
to the substrate, is perfectly coherent with the substrate, suggesting an
important strain, while the upper part shows a domain structure. The perfect
epitaxy between the film and the substrate can also be viewed on the
cross-section of Pr$_{0.5}$Ca$_{0.5}$MnO$_{3}$ deposited on SrTiO$_{3}$
(Fig.5). This film is grown in the [010]-direction, i.e. $2a_{P}$,
perpendicular to the substrate plane. The cross section along the
[110]-direction of the substrate clearly shows the perfect coherence of the
interface since the [100] or [001]-directions of the film match the
[110]-direction of the substrate. The lattice parameter length in this
direction is $a_{P}\sqrt{2}$.

In general, a strain is observed due to the epitaxial growth in very thin
films i.e. lattice parameters adopt those of the cubic lattice (see an
example of a compressive strain on SrTiO$_{3}$ \cite
{Razavi00-Thick,Zandbergen99-Thick}). In ultrathin films ($60$\AA ) of La$%
_{0.73}$Ca$_{0.27}$MnO$_{3}$ on\ SrTiO$_{3}$, the crystal structure imposed
by the substrate is different than the bulk \cite{Zandbergen99-Thick} and
leads to disorder effects \cite{Aarts98-Thick} or the formation of different
phases such as (La$_{0.7}$Sr$_{0.3}$)$_{3}$Mn$_{2}$O$_{7}$ in La$_{0.7}$Sr$%
_{0.3}$MnO$_{3}$ films \cite{Li00-Stu}. Microstructural studies also reveal
a slight distortion of the La$_{1-x}$Ca$_{x}$MnO$_{3}$ film, possibly
leading to a breakdown of the symmetry from orthorhombic to monoclinic (due
to the presence of spots in the electron pattern that are not allowed in the 
{\it Pnma}\ space group)\cite{Lebedev-PRB}: this suggests that the
structural situation might be different in thin film and in bulk material.
In contrast, Teodorescu {\it et al.} show that the structure and the
stoichiometry of the bulk target \ are perfectly reproduced in La$_{0.60}$Y$%
_{0.07}$Ca$_{0.33}$MnO$_{3-\delta }$ thin films \cite{Teodorescu00-Stu}.

A comparative study between La$_{2/3}$Ba$_{1/3}$MnO$_{3}$ (LBMO) and La$%
_{2/3}$Sr$_{1/3}$MnO$_{3}$ (LSMO) grown on SrTiO$_{3}$ shows that thick LBMO
presents a perfect epitaxy and grows coherently strained throughout the film
thickness, whereas the LSMO\ films are composed of two layers separated by
an intrinsic interface region containing a high density of defects \cite
{Wiedenhorst99-Sta}. Sometimes, secondary phases are observed \cite{Lu00-Cro}%
. In Pr$_{0.7}$Sr$_{0.3}$MnO$_{3}$ \cite{Mercey00-CM}, the deposition
temperature influences the microstructure and is thus, directly connected to
the $T_{C}$, which is depressed maybe due to the role of the grain
boundaries.

Returning to the structural characterization of CMR\ thin films, there are
roughly three tendencies that emerge from these studies. The first is that
the manganite films are much more sensitive to substrate-induced stress than
the analogous cuprate superconductors. The second deals with the presence of
two regimes of strain relaxation: one highly strained regime located close
to substrate and another above which is more relaxed. It is not clear
exactly where the interface \ is located or even if it exists in every film.
The last interesting point that has been shown by several groups is the
difference of crystal symmetry (lattice parameters, space group..) between
the thin film and the corresponding bulk material.

\subsection{Physical measurements}

The standard characterization of CMR\ thin films consists of resistance
measurements versus temperature in zero field and under an applied magnetic
field,\ by using the four probe technique and also in magnetization
measurements. Results pertaining to strain effects will be discussed in the
next section.

\subsubsection{Surface measurements}

Several groups focussed their studies on the surface \cite
{Choi99-Phys,Peng99-Phys,Dulli00-Phys}. Extensive thin film surface studies
were performed using two complementary techniques: Atomic Force Microscopy
(AFM) and Magnetic Force Microscopy (MFM) \cite
{Kwon97,Desfeux99-Phys,Soh00-Phys} .\ Work was mostly on La$_{1-x}$Sr$_{x}$%
MnO$_{3}$ since this material exhibits the highest Curie temperature. It was
also found that the properties of the surface are different from those of
the bulk in both the electronic and the composition \ point of view \cite
{Dulli00-Phys}. For example, the surface termination and the Ca surface
concentration depend on the overall Ca concentration in La$_{1-x}$Ca$_{x}$MnO%
$_{3}$ films \cite{Choi99-Phys} (the La/Ca ratio differs between the surface
and in the film). The surface of La$_{0.5}$Ca$_{0.5}$MnO$_{3}$ and La$%
_{0.66} $Ca$_{0.33}$MnO$_{3}$ show a highly ordered grain pattern induced by
strains \cite{Peng99-Phys}, and the La$_{0.65}$Sr$_{0.35}$MnO$_{3}$ surface
exhibits a surface phase transition at $240K$ (to be compared to $370K$ for
the bulk) \cite{Dulli00-Phys}.

\subsubsection{Transport across a grain boundary (GB)}

Grain boundaries\ (GB) strongly affect the properties of CMR materials. Low
field magnetoresistance (LFMR) has been reported and attributed to the
spin-dependent scattering of polarized electrons at the GB \cite{Gupta99-GB}%
. Researchers tried to enhance this property by artificially creating an
interface between two elements. We will describe here only the intrinsic
effect across natural GB\ (in polycrystalline thin films) and across
artificial GB (in films deposited on bicrystal substrates). The precise
influence of the substrate will be discussed in a separated part. Note that
another method has been utilized to create artificial GB by scratching the
LaAlO$_{3}$ substrate before the deposition of the LCMO film\cite
{Srinitiwarawong98-GB}. The MR is subsequent in a field of $2kOe$ and varies
with the field orientation with respect to the GB.

The simplest way of creating a natural GB is to grow the film on
polycrystalline substrates \cite{Gupta96-Lecoeur}. Most of this work was
done by {\it IBM}\cite{Gupta96-Lecoeur,Li97-Anis} on La-Ca-Mn-O (LCMO) and
La-Sr-Mn-O (LSMO) films. The $\rho -T$ curve of such films depends on the
grain size, as shown on Fig. 6: resistivity in zero field decreases when the
grain size increases, but the peak temperature of approximately $230K$ is
almost independent of the grain size\cite{Gupta96-Lecoeur}. Gu {\it et al.}
show that the Low Field Magnetoresistance at low temperature has a dramatic
dependence on the nature of the in-plane GB \cite{Gu98-Cro}. The reduction
of zero-field low-temperature resistivity is might be explained by the
spin-polarized tunneling accross half-metallic grains. Another possibility
to obtain polycrystalline samples is to decrease the deposition temperature.
The resulting GB results from a lower crystalline quality of the film \cite
{Teo98-GB}.

Bicrystal substrates having a single GB\ have also been used to study the
transport across a GB. LCMO and LSMO thin films were deposited on
bicrystalline SrTiO$_{3}$ substrates having a specific misorientation angle 
\cite{Mathur97-GB,Isaac98-GB}. To measure the properties of the GB\ only,
the film was patterned into a Wheatstone bridge. The GB\ resistance and its
magnetic field dependance is strongly dependent on the misorientation angle 
\cite{Isaac98-GB} (See Fig. 7). The MR increases with an increase of the
misorientation angle of the bicrystal \cite{Isaac98-GB}. The change of
resistance is $3\%$ under $2mT$ magnetic field at $300K$ for La$_{0.7}$Sr$%
_{0.3}$MnO$_{3}$. At $77K$, a large bridge resistance ($27\%$) is observed
during magnetic field sweeps between $\pm 200mT$ over a temperature range
down to $77K$. Steenbeck {\it et al.} utilized La$_{0.8}$Sr$_{0.2}$MnO$_{3}$
\ films grown on SrTiO$_{3}$ bicrystals with a misorientation angle of $36.8%
{{}^\circ}%
$ \cite{Steenbeck97-GB} or $24%
{{}^\circ}%
$ \cite{Steenbeck98-GB}. They found that the GB\ magnetoresistance occurs at
low temperature, separated from the intrinsic MR near $T_{C}$, and that the
sign of the MR at the GB\ depends on the domain structure and H \cite
{Steenbeck97-GB}. Moreover, current-voltage measurements show that the field
dependence might not be related to tunneling \cite{Todd99-GB}.

\subsubsection{Irradiation effects}

Irradiation, varying the ions dose\cite{Arora99-Phys,Ogale98-Phys}, was used
to look at the effect of columnar defects upon the thin films' properties 
\cite{Desfeux99-Phys,Arora99-Phys}. On irradiated La$_{0.7}$Sr$_{0.3}$MnO$%
_{3}$ samples, the MFM shows the existence of magnetic domains in different
magnetization directions, suggesting that the defects can be considered as
pinning centers for the magnetic domain walls \cite
{Desfeux99-Phys,Ogale98-Phys,Ogale00-Phys}. The effect of irradiation (with
90 Mev$^{16}$O) by was studied La$_{0.75}$Ca$_{0.25}$MnO$_{3}$. At a low
dose of $10^{11}$ $ions/cm^{2}$, the irradiation induces an increase of both
the Curie temperature and of the resistive temperature transition $T_{IM}$,
whereas for high doses a decrease is observed \cite{Bathe98-Phys} due to
enhancement of pinning for the magnetic domains walls. The film becomes
insulating and does not show any resistivity peak when the dose is higher
than $10^{14}$ $ions/cm^{2}$ (see Fig.8) \cite{Bathe98-Phys}. Irradiation
with $250MeV$\ Ag$^{17+}$ induces phase transformation in La$_{0.7}$Ca$%
_{0.3} $MnO$_{3}$ thin films \cite{Ogale00-Phys} indicating that the nature
of the ions plays also a role.

\subsubsection{Phase separation}

Phase separation was suspected \ in La$_{0.4}$Ba$_{0.1}$Ca$_{0.5}$MnO$_{3}$
films \cite{Hong00-Phys} and confirmed by\ the noise probe method \cite
{Raquet00-Phys} in La$_{2/3}$Ca$_{1/3}$MnO$_{3}$ films; in this work the
authors attribute the origin of the random telegraph noise to a dynamic
mixed-phase percolative process, where manganese clusters switched back and
forth between two phases that differ in their conductivity and
magnetization. This spatial inhomogeneity in doped manganite thin films was
also investigated in La$_{1-x}$Ca$_{x}$MnO$_{3}$ \cite{Fath99-Phys} using
Scanning Tunneling Microscopy. The phase separation is observed below the
Curie temperature where different structures of metallic and more insulating
areas coexist and are field dependent. This suggests that the insulator to
metal transition at $T_{C}$\ should be viewed as a percolation of metallic
ferromagnetic domains.

\subsubsection{Other experiments}

Magneto-optical measurements reveal the onset of the ferromagnetic
transition via the coercive field increase and the Kerr rotation \cite
{Bobo00-Phys}. Using this technique, the spontaneous formation of twins in La%
$_{2/3}$Ca$_{1/3}$MnO$_{3}$ films below $105K$ was also observed \cite
{Vlasko00-Phys}.

\section{Effects of Strains}

The CMR manganites are sensitive to all types of perturbations.\ In
particular, it has been shown in bulk that the internal (through the average
size of the A-site cation) or external pressure (via hydrostatic pressure)
can strongly influence the magnetotransport properties. Since the beginning
of the rediscovery of the CMR\ effect in Mn-based compounds, many studies
have been focussed on the strains in thin films.\ This is due to the fact
that Mn {\it eg} electrons, which determine most of the physical properties,
are coupled to the lattice degrees of freedom through the Jahn-Teller
trivalent manganese. Thus, strains affect the properties of the manganite
thin films, and, in consequence, one needs to correctly understand the
effects in order to obtain the desired properties.\ The following section
will discuss the two types of strains: in-the-plane (i.e. substrate-induced
strains, part A) and out-of-plane strain (i.e. thickness dependence, part B).

\subsection{Substrate-induced strains}

The first important parameter for successful thin film growth is undoubtedly
the substrate. For CMR\ materials, the same substrates as used for the HTSC
compounds were utilized. The most commonly used substrates to grow manganite
perovskites are MgO (cubic, $a=4.205$\AA ), SrTiO$_{3}$ (STO, cubic, $%
a=3.905 $\AA ), LaAlO$_{3}$ (LAO,\ pseudocubic, $a=3.788$\AA ), NdGaO$_{3}$
(NGO, orthorhombic with $a=5.426$\AA , $b=5.502$\AA\ and $c=7.706$\AA ) and
Si (cubic, $a=5.43$\AA ). Many authors have investigated the strain-effects
of the substrate by growing various films on different substrates \cite
{Wang00-Sta,Ranno99-Sta,Yeh97-Sta,Trtik99-Sta,Yu00-Cro}. They have
experimentally \cite
{Thomas98-Div23,Vlakhov98-Sta,Kwon97,Jin95,Lofland96,Tsui00-Thick,Ju98-Thick,Koo97-Anis}
or theoretically \cite{Millis98-Sta} studied the effect of strains on the
magnetoresistive properties of La$_{0.7}$Sr$_{0.3}$MnO$_{3}$ and La$_{0.7}$Ca%
$_{0.3}$MnO$_{3}$ or on the surface flatness \cite{Okawa00-Thick} for many
substrates. The physical properties of these materials depend on the overlap
between the manganese {\it d} orbitals and oxygen {\it p} orbitals, which
are closely related to the Mn-O-Mn bond angle and the Mn-O distance. As the
unit cell of the thin film is modified with respect to the bulk material,
the Mn-O distances and Mn-O-Mn angles are altered, inducing variations in
the electronic properties. We will review the main characteristics, such as
the structure and the physical properties, which are affected by the
substrate-induces strains.

\subsubsection{Modification of the structure and the microstructure}

The influence of the substrate upon the microstructure/structure, the
lattice parameters, the texturation and also the orientation of the film is
discussed in this section

Using Magnetic Force Microscopy, Kwon {\it et al.} \cite{Kwon97} showed on La%
$_{0.7}$Sr$_{0.3}$MnO$_{3}$ a ''feather-like'' pattern, indicating an
in-plane magnetization on (100)-SrTiO$_{3}$, while on (100)-LaAlO$_{3}$, a
''maze-like'' pattern corresponding to a perpendicular magnetization
anisotropy is seen. This kind of study was confirmed and extended recently
by Desfeux {\it et al.}\ on other substrates \cite{Desfeux01-APL}.

The influence of the substrate can principally be deduced from the lattice
parameters of the film and the effect of strain on lattice parameters has
been studied by many groups. These measurements lead to different
spin-structures. This is evident in La$_{1-x}$Sr$_{x}$MnO$_{3}$ for which
the spin-orbital phase diagram was obtained in the plane of strain-field ($%
c/a$ ratio) vs. doping {\it x} using the density-functional
electronic-calculation\cite{Konishi99-Thick}. The phase diagram of La$%
_{0.67} $Sr$_{0.33}$MnO$_{3}$ was also plotted for different substrates \cite
{Tsui00-Thick}.\ A strong dependence of anisotropy and Curie temperature on
lattice strain is observed. The effect of uniaxial strain was studied
theoretically by Ahn {\it et al. }\cite{Ahn}. Uniaxial strain produces
changes in the magnetic ground state, leading to dramatic changes in the
band structure and optical spectrum.

Both in-plane and out-of-plane lattice parameters are often modified by
stain effects when various substrates are used\cite
{Tsui00-Thick,Vlakhov98-Sta,Wu99-Thick,Ohshima00-Cro}. This is evidenced in
Fig.9 for $300$\AA\ thin films of Pr$_{0.7}$Sr$_{0.3}$MnO$_{3}$ (PSMO) grown
on LaAlO$_{3}$, SrTiO$_{3}$ and NdGaO$_{3}$ \cite{Wang00-Sta}. The 002 peaks
of the PSMO/LAO and PSMO/STO films are at 46.2%
${{}^\circ}$%
and 47.8%
${{}^\circ}$%
, corresponding to an out-of-plane parameter of 3.93 and 3.81 \AA ,
respectively (the diffraction peak of the PSMO film on NGO is almost
invisble from the substrate peak due to the small mismatch). These values
have to be compared to the lattice parameter close to 3.87\AA\ found in the
bulk PSMO. They indicate that the films are under tensile stress on STO
(decreasing in the growth direction and expanding in the plane) and under
compressive stress on LAO\ (decreasing in the plane and expanding in the
out-of-plane direction) due to the lattice mismatch between the film and the
substrate at room temperature (Fig.10). The strain effects\ on the
out-of-plane lattice parameter of La$_{0.7}$Ca$_{0.3}$MnO$_{3}$ (LCMO), are
enhanced with annealing \cite{Sun99-Thick}: the lattice expansion is 2-3
times larger in LCMO/STO than in LCMO/NGO. The stress also influences the
bond lengths and bond angles. Miniotas {\it et al.} have evaluated the Mn-O
and Mn-Mn distances in La$_{1-x}$Ca$_{x}$MnO$_{3}$ films grown by MBE.\ The
Mn-O bond length was found to be fixed at $1.975$\AA , independent of the
substrate types while the Mn-Mn distance (and subsequently the Mn-O-Mn bond
angle) was calculated to be $3.93$\AA\ for STO\ and $3.84$\AA\ for LAO \cite
{Miniotas01-Sta}.

Lattice mismatch (i.e. the difference of parameters between the film and the
substrate) influences not only the parameters of the film but also the
texturation (or epitaxy), i.e. the in-plane alignments. Usually, changes in
the in-plane orientation is observed only when the mismatch is small (e.g.
LCMO on LAO \cite{Gillman98-Cro} and is not realized on YSZ \cite
{Shreekala97-Div} or on MgO \cite{Vlakhov98-Sta}). Textured La$_{2/3}$Sr$%
_{1/3}$MnO$_{3}$ films were obtained on Si when buffer layers \cite
{Foncuberta99b,Fontcuberta99a} are used. La$_{0.7}$Ca$_{0.3}$MnO$_{3}$ was
grown using a buffer layer of CeO$_{2}$\cite{Zhang98-Cro} and LSMO\ with a
buffer of YSZ \cite{Trajanovic96-Cro} or a double layer Bi$_{4}$Ti$_{3}$O$%
_{12}$/SiO$_{2}$ \cite{Gu98-Cro,Gu97-Cro}. A highly conducting diffusion
barrier layer of TiN has also been utilized recently as a buffer layer \cite
{Kumar01-Cro}. This progress is interesting for technological reasons
especially when using Si substrates.

A surprising effect of lattice mismatch is related to the orientation of the
films, especially those that crystallizes in an orthorhombic perovskite
cell. This was first seen on YMnO$_{3}$ \cite{Savaldor98-Cro} which is
[010]-oriented on SrTiO$_{3}$, but [101]-oriented on NdGaO$_{3}$ or LaAlO$%
_{3}$. Similar results were obtained with Pr$_{0.5}$Ca$_{0.5}$MnO$_{3}$ \cite
{Prellier00-APL,Haghiri00-JAP}, Pr$_{0.7}$Sr$_{0.3}$MnO$_{3}$ \cite
{Mercey00-CM} or Pr$_{0.7}$Sr$_{0.3-x}$Ca$_{x}$MnO$_{3}$\cite{Mercey97-CM}.
It seems that this orientation can be generalized for every compounds that
crystallize in an orthorhombic structure. This dependence on the orientation
with the substrate is explained by the lattice mismatch ($\sigma $) which
should favor one orientation \cite{Prellier00-Thick}. Indeed, the mismatch
between the film and the substrate can be evaluated using the formula $%
\sigma =100\ast (a_{S}-a_{F})/a_{S}$ (where $a_{S}$ and $a_{F}$ respectively
refer to the lattice parameter of the substrate and the film). For Pr$_{0.5}$%
Ca$_{0.5}$MnO$_{3}$, the smaller mismatch on LaAlO$_{3}$ is obtained for the
[010]-axis in the plane ($\sigma _{LAO}=-0.4\%$), i.e. the [101]-axis
perpendicular to the substrate plane. In contrast, the smaller mismatch on
SrTiO$_{3}$ ($\sigma _{STO}=2.2\%$) is found for the [101]-axis is in the
plane and thus the [010]-axis normal to the surface of the substrate as
found experimentally \cite{Prellier00-Thick}.

\subsubsection{Influence on the physical properties}

Much work has been done on the influence of strain on the transport
properties, but we will also describe how the magnetic properties can be
changed. Many groups have focussed their studies on the modification of the
physical properties \cite{Krishnan96-Sta,Martin99-Cro} by the strain effects
since the most common properties such as the insulator-to-metal ($T_{IM}$)
transition and the Curie temperatures ($T_{C}$) are affected. In Fig.11, Koo 
{\it et al.} show this clear correlation between the substrate and the
physical properties for La$_{0.7}$Ca$_{0.3}$MnO$_{3}$ films: the $T_{IM}$\
and maximum MR shift to a higher temperature when changing a SrTiO$_{3}$ to
a LaAlO$_{3}$ substrate\cite{Koo97-Anis}.

Since the crystallinity of these films can be changed, as previously
discussed, Gillman {\it et al.} \cite{Gillman98-Cro} have prepared La$_{1-x}$%
Ca$_{x}$MnO$_{3}$ (x=0.41) films on substrates with different lattice
parameters by liquid delivery metalorganic chemical vapor deposition. Films
on LaAlO$_{3}$, closely lattice matched with the substrate, exhibit a high
degree of crystallization and a high magnetoresistance ratio as compared to
films deposited on Al$_{2}$O$_{3}$ or Y-ZrO$_{2}$. Similar results were
reported for La$_{0.8}$Sr$_{0.2}$MnO$_{3}$, which is epitaxial when grown on
(100)-LaAlO$_{3}$ and polycrystalline when grown and (100)-Si\cite
{Krishnan96-Sta}. Moreover, the $T_{IM}$ increases by $20K$\ when using
(011)-LaAlO$_{3}$ rather than (001)-LaAlO$_{3}$. Similar results have been
reported for La$_{2/3}$MnO$_{3-\delta }$ films grown on both Al$_{2}$O$_{3}$
and SrTiO$_{3}$ \cite{Chen98a-Cro} and also for La$_{0.67}$Ca$_{0.33}$MnO$%
_{3}$ \cite{Martin99-Cro}.\ The $T_{IM}$ is higher and the transition is
sharper for material grown on STO (300K) than on ALO (200K). Even if the $%
T_{IM}$ varies a lot, the Curie temperature is found to remain almost
constant, independent of the substrate, for many compounds such as La$_{2/3}$%
MnO$_{3-\delta }$ \cite{Chen98a-Cro} or La$_{0.7}$Sr$_{0.3}$MnO$_{3}$ \cite
{Haghiri00-Cro}.

The above evidence implies that the $T_{IM}$\ is directly connected to the
substrate. Often, the transition is at higher temperature and sharper when
the mismatch between the film and the substrate is smaller, probably due to
a high degree of epitaxy.

Strain not only influences the transport transitions, but the direction of
the magnetization as well (via lattice deformation): it is found to be
in-the-plane for films under tensile stress (fro example on SrTiO$_{3}$) and
out-of-plane for compressive stress (as in the case of LaAlO$_{3}$) \cite
{Kwon97,Haghiri00-Cro}. Using a wide-field Kerr microscope, magnetic domain
orientation and contrast of La$_{0.67}$Sr$_{0.33}$MnO$_{3}$/SrTiO$_{3}$
suggest a magnetic anisotropy \ with 
\mbox{$<$}%
110%
\mbox{$>$}%
easy axes \cite{Lecoeur97-Cro}. The easy direction is along [110] of the
pseudocubic unit cell , i.e. diagonal to the O-Mn-O bond direction for La$%
_{0.7}$Ca$_{0.3}$MnO$_{3}$ film grown on untwinned paramagnetic NdGaO$_{3}$
(001) \cite{Mathur01-Anis}.

The substrate-induced strain can also influence the optical properties, as
for La$_{0.67}$Ca$_{0.33}$MnO$_{3}$ \cite{Boris97-Phys,Vengalis00-Sta}. This
can be explained by the fact that the substrate-induced strain result in
modification in the Mn-O bonds and Mn-O-Mn bond angles and thus, in both the
corresponding phonon modes and electron-phonons interractions leading to
changes in the phonon frequencies and optical conductance. Note that the
strains can also induce a surface magnetization as for La$_{0.7}$Sr$_{0.3}$%
MnO$_{3}$ \cite{Lofland97-Div5}.

\subsubsection{Low Field Magnetoresistance (LFMR)}

The strain effects on the low field magnetoresistance\ (LFMR) was first
studied on polycrystalline La$_{0.67}$Sr$_{0.33}$MnO$_{3}$ \cite
{Gupta96-Lecoeur,Li97-Anis}. It was also extensively studied by Wang {\it et
al.}\ in Pr$_{0.67}$Sr$_{0.33}$MnO$_{3}$ \cite
{Wang00-Anis,Wang99-Anis,Wang98-Anis}. Films with compressive strains (on
LaAlO$_{3}$) show a large LFMR \cite{Wang99-Anis} when the field is applied
perpendicularly to the substrate plane (Fig.12), while they exhibit a
positive MR\ when the film is under tensile stress (on SrTiO$_{3}$) \cite
{Wang00-Anis}. Almost no LFMR is observed when the film is stress-free (on
NdGaO$_{3}$). O'Donnel {\it et al.} confirm that the LFMR\ depends on the
strains and the orientation of the field, by studying highly crystallized La$%
_{0.7}$Ca$_{0.3}$MnO$_{3}$ thin films made by Molecular Beam Epitaxy \cite
{O'Donnel96-Anis,O'Donnel00-Anis,O'Donnel97-Anis,Eckstein96-Anis,O'Donnel98-Thick}%
. It was also shown that the LFMR\ is dominated by the grain boundaries \cite
{Gupta96-Lecoeur,Li97-Anis} and its sign can be explained by a simple atomic 
{\it d}-state model \cite{Ziese98-Anis}. This idea of anisotropic MR \cite
{Wu99-Thick,Suzuki97-Thick,Suzuki99-Thick} was evidenced by La$_{0.7}$Sr$%
_{0.3}$MnO$_{3}$ deposited on (001)-SrTiO$_{3}$, (110)-SrTiO$_{3}$\cite
{Berndt00-Thick} and (110)-LaGaO$_{3}$. Magnetic anisotropy was also seen
recently on La$_{0.7}$Ca$_{0.3}$MnO$_{3}$ films grown on (001)-NdGaO$_{3}$ 
\cite{Mathur01-Anis}.

\subsubsection{Charge-ordered (CO) manganites}

Most of the experimental studies were done on manganites showing an
insulator-to-metal transition without field, but it was also shown that
substrate-induced strain can affect the properties of charge-ordered (CO)
compounds. CO\ is a phenomenon observed wherein electrons become localized
due to the ordering of heterovalent cations in two different sublattices ($%
Mn^{3+}$ and $Mn^{4+}$). The material becomes insulating below the CO\
transition temperature, but it is possible to destroy this state and render
the material metallic by, for example, the application of a magnetic field 
\cite{Rao00-Rev} but, an electric field can also induced \ insulator-metal
transitions in thin films of CO manganites\cite{Ponmambalam99-CO}. Pr$_{0.5}$%
Ca$_{0.5}$MnO$_{3}$ is an example of such a compound.\ In this case, a
tensile stress can decrease the melting magnetic field \cite{Prellier00-APL}
whereas a compressive strain induces a locking of the structure\cite
{Haghiri00-JAP} at low temperature (i.e. under cooling, when the in-plane
parameters of the film are equal to the parameters of the substrate, they
are kept at this value). This idea of locking was confirmed in Pr$_{0.5}$Sr$%
_{0.5}$MnO$_{3}$ where the structural and physical transitions are
suppressed under cooling\cite{Mercey01-APL}, as compared to the bulk (note
that, even if the compound Pr$_{0.5}$Sr$_{0.5}$MnO$_{3}$ is not a typical
CO, it has some similarities in the physical properties). In this material,
the A-type antiferromagnetic phase with the {\it Fmmm} structure, which is
obtained at low temperature (below 135K) in the corresponding bulk compound,
is not observed in the thin film. The consequence of the absence of
structural transitions is that magneto-transport properties are affected.
There is no antiferromagnetism (i.e. the A-phase) at low temperature. The
material only becomes ferromagnetic insulating. This is one of the very few
examples of substrate-induced strain upon the film structure. These results
show that the strain effect can destabilized the charge-ordered state for CO
materials but, surprisingly, it seems also possible to induce a CO state
when the film composition is not a CO\ type (i.e. if the film has an
insulator-to-metal transition without the presence of a magnetic field).
This has been shown in a normally metallic La$_{0.7}$Ca$_{0.3}$MnO$_{3}$
compound where the lattice-mismatch strain effects leads to a strain-induced
insulating state \cite{Biswas01-PRB}.\ This insulating behavior is related
to the coexistence of a metallic state with a possibly charge-ordered
insulating state\cite{Biswas01-PRB,Biswas00-Thick}.

\subsection{Thickness dependence}

\subsubsection{Lattice parameters}

The influence of the thickness ($t$) is primary seen upon the lattice
parameters of the films (in-plane and out-of-plane parameters). Usually, the
volume of the unit cell is conserved in the thin film as compared to the
bulk. In order to verify this result, the evolution of the three-dimensional
strain states and on \ crystallographic domain structures was studied on
epitaxial La$_{0.8}$Ca$_{0.2}$MnO$_{3}$ as a function of lattice mismatch
with two types of (001)-substrates, SrTiO$_{3}$ and LaAlO$_{3}$\cite
{Rao98-Sta,Nath99-Sta}. Surprisingly, it was shown, using normal and grazing
incidence x-ray diffraction techniques, that the unit cell volume is not
conserved and varies with the substrate as well as the film thickness (Fig.
13).

But the main result is that for a tensiled film (under expansion in the
plane of the substrate), the out-of plane and in-plane parameters gradually
increases and decreases, respectively, as a function of the film thickness 
\cite{Prellier00-Thick}. For example, in the case of Nd$_{2/3}$Sr$_{1/3}$MnO$%
_{3}$ grown\ on SrTiO$_{3}$, the out-of-plane parameter increases from a
value of $3.8$\AA\ for a $200$\AA\ thick film to $3.86$\AA\ for a $1000$\AA\
film, which is close to the bulk value (Fig. 14)\cite{Barman00-Thick}. The
scenario is the opposite when the film is compressively strained as in La$%
_{0.7}$Sr$_{0.3}$MnO$_{3}$ on (100)-LaAlO$_{3}$ \cite{Suzuki00-Sta}: the
out-of-plane parameter decreases from 3.94\AA\ for a 300\AA\ thick film to
3.9\AA\ for a 4500\AA\ thick film, while at the same time the in-plane
parameter changes from 3.82\AA\ to 3.88\AA . The film is not completely
relaxed until it reaches a thickness on the order of 1000\AA . In Nd$_{0.5}$%
Sr$_{0.5}$MnO$_{3}$ deposited on (001)-LaAlO$_{3}$ \cite{Prellier99-Thick},
two regimes were observed using XRD: one which was strained (close to the
substrate), and a quasi-relaxed component in the upper part of the film, the
latter increasing with film thickness.\ 

As previously reported, increasing film thickness leads to a change of the
symmetry of the film. This was systematically studied by looking at various
film compositions vs. thickness by Yu {\it et al.\cite{Yu00-Cro}. }They
found a strong tetragonal lattice strain using HREM\ and XRD\
characterization. This is more important for a composition in which the bulk
structure is orthorhombic, as in (La$_{1-x}$Pr$_{x}$)$_{0.7}$Ca$_{0.3}$MnO$%
_{3}$, as compared to La$_{1-x}$Na$_{x}$MnO$_{3}$ or La$_{0.7}$Sr$_{0.3}$MnO$%
_{3}$ where the structure is rhombohedral\cite{Yu00-Cro}.

\subsubsection{Physical properties}

The physical properties of manganites\cite
{Okawa00-Thick,Sun99-Thick,Sirena00-Thick} such as insulator-to-metal
transition \cite{Walter00-Sta,Steren00b-Thick}, magnetoresistance \cite
{Jin95}, coercive field \cite{Steren00a-Thick} or microstructure \cite
{Aarts98-Thick,Gross99-Thick,Prauss99-Sta} are strongly dependent on the
thickness. As an example, the MR\ value calculated as ($\Delta R/R(H)$ for $%
H=6T$) exhibits a strong dependence on film thickness as shown in Fig. 15
for La$_{0.7}$Ca$_{0.3}$MnO$_{3}$.\ The curves show a maximum MR for a
thickness near 1100\AA\ with a value of $10^{6}\%$ and, on either side of
the peak, the MR ratio is drastically lower. Transport properties are mostly
affected and the magnetization is\ only moderately changed \cite
{Sirena00-Thick} with thickness as seen for La$_{0.6}$Sr$_{0.4}$MnO$_{3}$
deposited on MgO or SrTiO$_{3}$. At an intermediate thickness around $1000$%
\AA , the films usually recover the properties of the bulk compounds. Even
when the film is under low epitaxial stress (case of NGO), $T_{IM}$ varies
greatly from $182K$ ($35$\AA ) to $264K$ ($1650$\AA ), as seen in Fig. 16
for La$_{0.7}$Ca$_{0.3}$MnO$_{3}$ \cite{Walter00-Sta}.

Films thinner than 1000\AA\ have properties different from the bulk and are
most of the time unusual. For example thin La$_{1-x}$Ba$_{x}$MnO$_{3}$ films
(t%
\mbox{$<$}%
1000\AA ) exhibits a $T_{C}$ higher than in the bulk due to an anomalous
tensile strain effect when deposited on SrTiO$_{3}$. Consequently, the
resulting film shows room temperature ferromagnetism and an enhancement of
the magnetoresistance\cite{Kanki00-Cro}. In (La,Ca)MnO$_{3}$ films, the
thinnest films which present full magnetization, grow with the {\it b} axis
of the structure perpendicular to the substrate, whereas the thicker films
grow with the {\it b}-axis in the plane of the substrate and do not present
full magnetization \cite{Aarts98-Thick}. Another example of the thickness
dependence is seen in La$_{0.67}$Ca$_{0.33}$MnO$_{3}$ on SrTiO$_{3}$, which
is ferromagnetic around 150K but remains insulating \cite{Zandbergen99-Thick}%
. Biswas {\it et al.} \cite{Biswas00-Thick} have explained this behavior by
the coexistence of two different phases, a metallic ferromagnet (in the
highly strained region) and an insulating antiferromagnet (in the low strain
one). This nonuniformity induces, under a magnetic field, an
insulator-to-metal transition resulting in a large CMR effect. But, the
metallic behavior of the bulk La$_{0.7}$Ca$_{0.3}$MnO$_{3}$ can be retained
for a thickness down to 60\AA\ when the SrTiO$_{3}$ substrate is treated to
obtain an atomically flat TiO$_{2}$ terminated surface \cite{Padhan01-Sta}.
For La$_{0.9}$Sr$_{0.1}$MnO$_{3}$ (t%
\mbox{$<$}%
50nm) on (100)-SrTiO$_{3}$, Razavi {\it et al.} \cite{Razavi00-Thick}
reported an unexpected insulator-to-metal transition, most probably due to
La-deficiency.\ Nevertheless, Sun {\it et al. }\cite{Sun99-Thick} have
estimated the ''dead-layer'' for La$_{0.67}$Sr$_{0.33}$MnO$_{3}$ to be
around $30$\AA\ for NdGaO$_{3}$ and $50$\AA\ for LaAlO$_{3}$ (Fig. 17). The
magnetic, transport and structural properties of La$_{0.7}$Sr$_{0.3}$MnO$%
_{3} $ deposited were MgO was interpreted recently in terms of a magnetic ($%
10$\AA ) and an electrical (insulating) dead layer (respectively of $10$\AA\
and $4$\AA\ thick) \cite{Borges01-Thick}.

More recently, the robustness of the charge-ordered (CO) state was studied
by Prellier {\it et al.} \cite{Prellier00-Thick}. In Pr$_{0.5}$Ca$_{0.5}$MnO$%
_{3}$, the thicker film induced the less stable state, i.e. a small magnetic
field as compared to the bulk is required to destroy the CO\ state and
induce a metallic behavior (Fig.18). In Nd$_{0.5}$Sr$_{0.5}$MnO$_{3}$, the
(110)-films show a strained and a quasi-relaxed layer. The latter increases
with film thickness whereas the strained one has a constant thickness\cite
{Prellier99-Thick}. The coexistence of two strain regimes inside the same
film was also seen in La$_{0.66}$Ca$_{0.33}$MnO$_{3}$ films on SrTiO$_{3}$ 
\cite{Lebedev-PRB} and LaMnO$_{3}$ deposited on NdGaO$_{3}$ \cite
{Mercey99-Sta}: at the interface a cubic-like dense layer ($50$\AA\ thick)
is observed while the upper layer shows a columnar growth. These two
distinct thickness ranges behave differently with respect to the
thickness-dependence of the magnetotransport properties \cite{Wang00-Sta};
the upper range ($t>200$\AA ) is weakly thickness-dependent whereas the
lower one not. These results \cite{Wang00-Sta,Konishi99-Thick} show the
evidence for the effect of Jahn-Teller type distortion and confirm
theoretical explanations \cite{Millis98-Sta}. In Nd$_{2/3}$Sr$_{1/3}$MnO$%
_{3} $ films, the release of the strain as the thickness increases \cite
{Barman00-Thick} results in a first-order phase transition.

Thus, these results show the thickness-dependence of the physical properties
of the film, but it seems difficult to estimate these changes precisely. For
example, considering a Pr$_{0.7}$Sr$_{0.3}$MnO$_{3}$ film deposited on NdGaO$%
_{3}$, is it possible to evaluate the $T_{IM}$\ and the $T_{C}$\ for
2000\AA\ thick film ? There is no report of such calculations, and one of
the reasons is that the properties of the film depend not only of the
substrate, but also on the growth conditions. It will be necessary to answer
this question in the future.

\section{Potential of thin film growth: the design.}

The improvement in controlled heterostructures and multilayers is a
necessary stage for the realization of many devices and circuits. Structures
with new properties such as superlattices were also widely studied.

\subsection{New structures}

Thin film methods offer a powerful and versatile technique for growing new
structures, as previously seen for example in cuprates. This is due to
strain effects that can stabilize structures which do not exist under
classical conditions of pressure and temperature. For example, various
metastable perovskites, which can not be formed in bulk or can only be
prepared under high pressure, such as BiMnO$_{3}$ \cite{Ohshima00-Cro}, YMnO$%
_{3}$ \cite{Savaldor98-Cro}, atomically ordered LaFe$_{0.5}$Mn$_{0.5}$O$_{3}$
\cite{Ueda01-Cro} are synthesized via a pulsed laser method or by using
injection MOCVD like NdMn$_{7}$O$_{12}$\cite{Bosak00-Cro}.

Also interesting is the construction of new compounds, such as artificial
superlattices, that show unique physical properties since different types of
magnetism can be combined by building the desirable structure at the atomic
layer level \cite{Tabata97-SL}. Growth conditions such as the oxygen
pressure or the deposition temperature are easy to control in the thin film
process, allowing the synthesis of metastable phases.

Another approach to obtain exotic properties via new phases is the method of
artificial superlattices. Preliminary films were grown by stacking a
magnetic layer (La$_{0.7}$A$_{0.3}$MnO$_{3}$ with A=Sr, Ba...) and another
perovskite, usually insulator (such as SrTiO$_{3}$) \cite{Kwon97-SL}. This
allows a continuous variation of the in-plane coherency strain in the films 
\cite{Lu00-SL,Jo99-SL,Izumi00-SL}. High quality films showing a clear
chemical modulation \ by the presence of satellite peaks around the main
diffraction peak were obtained \cite{Lu00-SL,Luo00-SL}. In the case of La$%
_{0.7}$Ca$_{0.3}$MnO$_{3}$ (LCMO), the metallic transition is suppressed and
the MR\ enhanced at low temperature when the thickness of the LCMO layer is
decreased to $25$\AA\ \cite{Jo99-SL}. The MR ($MR=100\ast \lbrack
R(0)-R(H)]/R(0)$) is calculated to be $85\%$ at $H=5T$ \ over a wide
temperature range ($10-150K$) (Fig. 19). A systematic study of La$_{0.7}$Ba$%
_{0.3}$MnO$_{3}$/SrTiO$_{3}$ superlattices shows that the decrease of the La$%
_{0.7}$Ba$_{0.3}$MnO$_{3}$ (LBMO) layer thickness results in the broadening
of the MR\ peak vs. temperature\cite{Kwon97-SL}. Such studies also confirm
the importance of strains and the relevance of Jahn-Teller electron-phonon
coupling in doped manganites, as pointed out by Lu {\it et al.} \cite
{Lu00-SL}. Following the same idea, La$_{2/3}$Ba$_{1/3}$MnO$_{3}$/LaNiO$_{3}$
multilayers were synthesized. Magnetization measurements show evidence of
antiferromagnetic coupling between LBMO layers when the thickness of the
LaNiO$_{3}$ spacer is $15$\AA\ or less\cite{Nikolaev99-SL}.

The magnetic exchange interactions have been extensively studied in La(Sr)MnO%
$_{3}$/LaMO$_{3}$ (M=Fe, Cr, Co, Ni) \cite{Tanaka99-SL,Tanaka00-SL}. The
authors showed that the ferromagnetism is systematically affected by the
adjacent magnetic layers via the interface, and, they propose an expression
of T$_{C}$ on the basis of the molecular field image. The magnetotransport
properties of superlattices such as La$_{0.6}$Pb$_{0.4}$MnO$_{3}$/La$_{0.85}$%
MnO$_{3}$\cite{Sahana99-SL} or La$_{0.7}$MnO$_{3}$/Pb$_{0.65}$Ba$_{0.05}$Ca$%
_{0.3}$MnO$_{3}$\cite{Pietambaram01-SL} were also investigated.\ An
enhancement of the magnetoresistance is obtained in these materials. {\it c}%
-axis YBa$_{2}$Cu$_{3}$O$_{7}$/La$_{0.67}$Ba$_{0.33}$MnO$_{3}$ superlattices
were also grown \cite{Jakob95-SL}. Above $T_{C}$, the CMR\ persists up to
room temperature, and below $T_{C}$ the superlattices exhibit a
quasi-two-dimensional superconductivity of the YBa$_{2}$Cu$_{3}$O$_{7}$\
layers coexisting with magnetism in the La$_{0.67}$Ba$_{0.33}$MnO$_{3}$ \cite
{Jakob95-SL}. \ An increase in the thickness of the antiferromagnetic La$%
_{0.6}$Sr$_{0.4}$FeO$_{3}$ layer in between La$_{0.6}$Sr$_{0.4}$MnO$_{3}$
layers induces a strong magnetic frustration around the superlattice
interfaces, leading to a reduction of the magnetic temperature transition
and of the ferromagnetic volume \cite{Izumi00-SL}.

Salvador{\it \ et al. }used the PLD\ technique to create A-site ordering in
films of (LaMnO$_{3}$)/(SrMnO$_{3}$) superlattices \cite{Salvador99-SL}. An
increase of the superlattice period leads to a decrease in the $T_{C}$ and
in $T_{IM}$ \ or in a low magnetization value.

\subsection{Some Devices}

The intense efforts of the condensed matter community in the area of CMR
thin films have led to a more precise understanding of the growth of thin
film oxides even if the utilization of the materials into devices has not
proven to be viable yet. The aim of this article is to give an overview of
the manganite thin films and, for this reason, we will not go into details
of the realization of devices. A more comprehensive description of the
devices made using CMR materials can be found elsewhere \cite{Venky98-Div23}%
. Examples of devices include magnetic field sensors, electric field devices 
\cite{Mathews97-APP}, uncooled infrared bolometers\cite{Goyal97-Liv} or low
temperature hybrid HTSC-CMR devices \cite{Goldman99-APP}. Some of these are
briefly discussed below.

A magnetic tunnel junction is a structure composed of two ferromagnetic (FM)
layers separated by an insulator barrier (I) and have attracted attention
due to their properties of tunneling magnetoresistance (TMR). However, to
obtain TMR (FM/I/FM junction) with 100\% efficiency, it is necessary to have
a perfect half-metal (i.e. a 100\% spin polarization). Such a property was
confirmed by spin resolved photoemission measurements in the case of the La$%
_{0.7}$Sr$_{0.3}$MnO$_{3}$ (LSMO) compound \cite{Park98-APP}. Junctions with
LSMO which show a TMR\ effect\cite
{Kwon98-APP,Sun96-APP,Viret97-APP,Ghosh98-APP,Ogale98-APP,Deteresa99-APP}. A
large MR\ in $83\%$ at a low field of $10$Oe at $4.2K$ in a trilayer film of
LSMO/STO/LSMO was observed\cite{Li97-APP} (Fig. 20). Note that the top layer
can also be Co \cite{Deteresa99-APP}, half-filled ferrimagnetic Fe$_{3}$O$%
_{4}$ \cite{Ghosh98-APP,Ogale98-APP} or La$_{0.7}$Sr$_{0.3}$MnO$_{3}$ \cite
{Kwon98-APP,Sun96-APP}. Using Fe$_{3}$O$_{4}$, a positive MR is observed
which could be attributed to the inverse correlation between the
orientations of the carrier spins in the two\ FM\ layers \cite{Ghosh98-APP}.

The electric field effect has been investigated, in which the top layer can
be paramagnetic, such as STO \cite{Ogale96-APP}, or ferroelectric layer,
such as PZT\ (PbZr$_{0.2}$Ti$_{0.8}$O$_{3}$\cite{Mathews97-APP}), and\ the
bottom layer is a CMR material. but the changes are more profond in the case
of PZT where only $3\%$ change in the channel resistance is measured over a
period of 45min at room temperature which makes this attractive for
nonvolatile ferroelectric field effect devices \cite{Mathews97-APP}.

The large temperature coefficient of resistance ($TCR$, calculated as $%
(1/R)(dR/dT)$) just below the resistivity peak makes these CMR materials
interesting for bolometric detectors \cite{Rajeswari96-APP,Lisauskas00-APP}.
However, for a given material the TCR\ decreases as $T_{C}$ or $T_{IM}$
increase \cite{Goyal97-Liv}. A\ $TCR$\ of $7\%/K$ is obtained for LCMO at $%
250K$.

Hybrid structures consisting of HTSC-CMR have also been made for use in spin
injection devices \cite{Spin97,Dong97-APP,Spin99}.

\section{Conclusion}

The structural, magnetic and transport properties of manganite thin films
have been presented in this article. As seen, the colossal magnetoresistive
oxides display an exciting diversity of behavior in the form of thin films,
and an extremely large amount of work has been carried out on thin films
showing the great potential of their magnetic and transport properties. It
has been shown that the structural and physical properties of these oxides
are strongly dependent on the deposition procedure, chemical composition and
applied strain. For this reason, the direct comparison of data between a
thin film and a bulk material (ceramic or single crystal) is difficult due
to the stress in the thin film.

It has also been shown that devices are interesting and potentially useful
for magnetic sensors. Prior to the fabrication of such devices, it will be
necessary to characterize the materials more comprehensively, in particular
from the view point of the structure and the microstructure. This is clearly
evidenced by the fact that intrinsic phenomena such as the substrate-induced
strain and the thickness dependence, which are directly related to the thin
film process, strongly affect the magnetotransport properties. These results
suggest that the local lattice distortions of the Mn-O bonds in the
manganite thin films contribute to changes in the physical properties.

There are two main ideas that should be considered in the future based on
the recent results. It is now recognized that the strains directly affect
the lattice parameters. In addition, researchers have noted that there is a
clear relation between the oxygen content (or indirectly the Mn$^{3+}$/Mn$%
^{4+}$ ratio) and the lattice parameters of the unit cell of the film. Thus,
one should ask the following question: what is the relation between the
oxygen content and strain? This triangular connection must be investigated
precisely and explained in the future. The second main direction is related
to the stress, because despite the large amount of work published on
manganite thin films, there is still no direct proof of substrate-induced
strains: researchers have only found indirect correlations at room
temperature. More sophisticated mechanisms going beyond classical concepts
(i.e. by looking at the evolution of the structure under cooling) and
theoretical work, in particular by quantifying the stress for these oxide
films, are required to understand the nature of this class of compounds.

\bigskip

Acknowledgments:

We would like to acknowledge Dr. A. Maignan, Dr. A. Ambrosini, Prof. B.
Raveau (Laboratoire CRISMAT, Universit\'{e} de Caen), Dr. L. Mechin
(Laboratoire GREYC, ISMRA-Universit\'{e} de Caen), Dr. A. Anane (Unit\'{e}
mixte CNRS/Thales, Orsay), Dr. A.M. Haghiri-Gosnet (IEF, Universit\'{e}
d'Orsay), Prof. R.L. Greene (Center for Superconductivity Research,
University of Maryland), Prof. P.A. Salvador (Carnegie Mellon, University of
Pittsburg) and Dr. R. Desfeux (Universit\'{e} d'Artois) for fruitful
discussions and careful reading this article. We also thank M.\ Morin for
helping with the preparation of the manuscript.

Fig. 1: Typical experimental setup of a Pulsed Laser Deposition system used
for growing oxides.

Fig. 2: $\rho (T)$ and MR\ ratio of La$_{1-x}$Sr$_{x}$MnO$_{3}$ thin films
is zero field and a field of 5T for as-grown (top) and annealed at 950$%
{{}^\circ}%
C$ for 10h in pure N$_{2}$ gas (bottom). (a): x=0.33, (b): x=0.2 and (c):
x=0.16 (reproduced from ref. 98).

Fig. 3: $M(T)$ for as-deposited and post-annealed La$_{0.8}$MnO$_{3-\delta }$
films. Annealing 1-600$%
{{}^\circ}%
C,3h$; Annealing 2-600$%
{{}^\circ}%
C,24h;$ Annealing 3-700$%
{{}^\circ}%
C,3h;$ Annealing 4-800$%
{{}^\circ}%
C,3h$ (Reproduced from Ref. 99).

Fig. 4: Map of the reaction conditions for different manganite thin films on
SrTiO$_{3}$.\ The growth conditions for the {\it c}-axis oriented (La,Sr)$%
_{3}$Mn$_{2}$O$_{7}$ films (type 3) is restricted to the hatched region.
Detailed structure of films with type 1 (open circles) and type 2 (open
triangles) are not identified (Reproduced from Ref.\ 106).

Fig. 5: [010]-cross-section of Pr$_{0.5}$Ca$_{0.5}$MnO$_{3}$ deposited on
SrTiO$_{3}$ taken at room temperature.

Fig. 6: $\rho $(T) under different magnetic fields for polycrystalline La$%
_{0.7}$Ca$_{0.3}$MnO$_{3}$ (LCMO) films with different grain sizes and an
epitaxial film. The inset shows the zero-field resistivity at 10K as a
function of the average grain size for and La$_{0.75}$MnO$_{3}$ (LXMO)
(Reproduced from Ref. 133).

Fig. 7: Normalized GB\ MR\ as a function of the applied field for different
bicrystal angles measured at room temperature. The inset shows the (a) low
and (b) high field dependences for the 24%
${{}^\circ}$%
device. The applied magnetic field is in plane and perpendicular to the GB
(Reproduced from Ref. 138).

Fig. 8: $\rho $(T) for as-deposited films and irradiated films at different
ion doses (Reproduced from Ref. 147).

Fig. 9: XRD\ pattern in the range 45-50$%
{{}^\circ}%
$ of 300\AA\ thick Pr$_{0.67}$Sr$_{0.33}$MnO$_{3}$ films on LaAlO$_{3}$,
NdGaO$_{3}$ and SrTiO$_{3}$. The arrows indicate the 002 peaks of Pr$_{0.67}$%
Sr$_{0.33}$MnO$_{3}$ (Reproduced from Ref. 63).

Fig. 10: Schematic structure of a film grown under tensile and compressive
stress in the plane. Note the compression or the elongation of the
out-of-plane parameter depending on the nature of the stress.

Fig. 11: \ (a) $\rho $(T) of La$_{0.7}$Ca$_{0.3}$MnO$_{3}$ on various
substrates under 0 and 1T magnetic field. (b) MR defined as ($\rho (0)-\rho
(1T))/\rho (0)$) normalized to the value at T=317K (Reproduced from Ref.
159).

Fig. 12: Normalized R(H) curves of 75$\ $\AA\ thick Pr$_{0.7}$Sr$_{0.3}$MnO$%
_{3}$ films on LaAlO$_{3}$ measured with the field applied parallel (H//)
and perpendicular (H$\perp $) to the plane of the substrate. In the H$\perp $
geometry, a very sharp

hysteresis loop and large MR ratio are observed. Arrows indicate the
scanning sequence of the magnetic field (Reproduced from Ref. 183).

Fig. 13: Thickness dependence of the perovskite unit cell volume of
epitaxial La$_{0.8}$Ca$_{0.2}$MnO$_{3}$ films on (001)-LaAlO$_{3}$
(triangles) and (001)-SrTiO$_{3}$ (circles). Large deviations of the lattice
parameters from those of the bulk are observed. As film thickness increases,
both in-plane and out-of-plane lattice parameters tend to deviate away from
those of the substrates towards the bulk value (Reproduced from Ref.\ 196).

Fig. 14: {\it c}-axis lattice parameter of Nd$_{0.7}$Sr$_{0.3}$MnO$_{3}$
thin films deposited on SrTiO$_{3}$\ as a function of the thickness
(Reproduced from Ref. 198).

Fig. 15: Thickness dependence of the MR \ for La$_{0.67}$Ca$_{0.33}$MnO$_{3}$
thin films grown on (100)-LaAlO$_{3}$ (Reproduced from Ref.155).

Fig. 16: \ $\rho $(T) of La$_{0.7}$Ca$_{0.3}$MnO$_{3}$ grown on (110)-NdGaO$%
_{3}$ vs. film thickness (Reproduced from Ref. 201).

Fig. 17: (a) Temperature dependence of the resistivity for La$_{0.67}$Sr$%
_{0.33}$MnO$_{3}$ films with varying thickness on NdGaO$_{3}$ and LaAlO$_{3}$%
. (b): Thickness dependence of the conductance of films at 14K (Reproduced
from Ref. 166).

Fig. 18: $\rho $(T) under varying magnetic fields for different thickness of
Pr$_{0.5}$Ca$_{0.5}$MnO$_{3}$ thin films grown on SrTiO$_{3}$. Arrows
indicate the direction of the temperature dependence (Reproduced from Ref.
174).

Fig. 19: MR (MR=100$\times $[R(0T)-R(5T)]/R(0T)) vs. temperature for a LCMO
(55\AA )/STO (160\AA ) superlattice (20 periods) and a single\ LCMO layer.
Note the broadening of the MR\ peak and the MR\ of 85\% at 5T from 10K to
150K (Reproduced from Ref. 212).

Fig. 20: MR vs. applied magnetic field at different temperatures for a
tunnel junction with a rectangular $2.5\times 12.5\mu m$ top electrode. The
moments of both electrodes are shown at various fields. Magnetic field is
applied is applied along the easy axis of the rectangle (see inset)
(Reproduced from Ref. 234).

\end{document}